\documentclass[
reprint,
superscriptaddress,
 amsmath,amssymb,
 aps,
prb,
]{revtex4-2}

\usepackage{dcolumn}
\usepackage{graphicx}
\usepackage{hyperref}
\usepackage{color}
\usepackage{tikz}
\usepackage{ascmac,amsmath,amsfonts,amsthm,amssymb}
\usepackage{url}

\newenvironment{itmbx}[1]{\begin{itembox}[l]{#1}}{\end{itembox}}

\newcommand{\ket}[1]{| #1 \rangle}
\newcommand{\bra}[1]{\langle #1 |}

\begin{document}

\preprint{APS/123-QED}

\title{Topological charge pumping in quasiperiodic systems characterized by Bott index}

\author{Mao~Yoshii}
 \email{mao@g.ecc.u-tokyo.ac.jp}
\affiliation{Department of Applied Physics, The University of Tokyo, Hongo, Tokyo, 113-8656, Japan}
\author{Sota~Kitamura}
\affiliation{Department of Applied Physics, The University of Tokyo, Hongo, Tokyo, 113-8656, Japan}
\author{Takahiro~Morimoto}
\affiliation{Department of Applied Physics, The University of Tokyo, Hongo, Tokyo, 113-8656, Japan}
\affiliation{JST, PRESTO, Kawaguchi, Saitama, 332-0012, Japan}
\date{\today}

\begin{abstract}
   We study topological charge pumping in one-dimensional quasiperiodic systems.
   Since these systems lack periodicity, we cannot use the conventional approach based on the topological Chern number defined in the momentum space.
   Here, we develop a general formalism based on a real space picture using the so-called Bott index.
   We extend the Bott index that was previously used to characterize quantum Hall effects in quasiperiodic systems, and apply it to topological charge pumping in quasiperiodic systems. The Bott index allows us to systematically compute topological indices of charge pumping, regardless of the detail of quasiperiodic models. We apply this formalism to the Fibonacci-Rice-Mele model which we made from Fibonacci lattice, a well-known quasiperiodic system, and Rice-Mele model. We find that these quasiperiodic systems show topological charge pumping with a multi-level behavior due to the fractal nature of the Fibonacci lattice.
   Such multi-level pumping behaviors can be understood by a real space renormalization group analysis.
\end{abstract}

\maketitle

\section{Introduction}
Topology plays a central role in recent studies of quantum materials \cite{RevModPhys.82.3045,RevModPhys.83.1057,RevModPhys.88.035005}. Topological phases of matter arise from the nontrivial topology of electron wave functions in crystals, and exhibit characteristic quantized response phenomena. 
Quantum Hall effect (QHE) is the canonical example where the Hall conductivity shows quantization into the Chern number \cite{PhysRevLett.49.405}. The Chern number is a topological quantity consisting of Berry curvature of Bloch wave functions that quantifies the nontrivial geometry of momentum space. 
Topological charge pumping in one dimension is closely related to the quantum Hall effect in a sense that it is also characterized by the Chern number and the Berry curvature \cite{resta-RMP,vanderbilt-kingsmith}. In the case of charge pumping, the corresponding Berry curvature measures nontrivial geometry in the two-dimensional space spanned by the momentum and the pumping parameter. 

Quasiperiodic systems are the systems that possess long-range order without the translational symmetry. An early example of quasiperiodic structure is discovered in the system of alloys \cite{PhysRevLett.53.1951}, and quasiperiodicity has later been found in various systems \cite{PhysRevLett.109.116404,vardeny2013optics,kamiya2018discovery,tsai2000stable,collins2017imaging}. The structure of quasiperiodic crystals can be regarded as a projection of higher-dimensional-crystalline structure \cite{DEBRUIJN198139,DEBRUIJN198153}, and would allow us to access the physics of higher-dimensional-space that is usually inaccessible in three-dimensional crystals. 
Recently, van der Waals (vdW) heterostructure of two dimensional thin films has been realized and intensively studied, including twisted bilayer graphenes  \cite{Bistritzer12233,cao2018correlated,PhysRevB.99.165430} and interface of transition metal dichalcogenides  \cite{wang2020correlated,Akamatsu68}.
VdW heterostructures made of different crystals can be also considered as quasiperiodic systems \cite{Akamatsu68,kennes2021moire}, which provides an interesting platform for quasiperiodic structures due to their controllability and a rich variety of material combinations. 

Topology and geometry in quasiperiodic systems are an interesting subject. The conventional characterization of topological phases relies on the momentum space that requires translation symmetry and is not directly applicable to quasiperiodic systems that lacks translation symmetry. Therefore, an alternative description for topological phases is required for quasiperiodic systems. Indeed, several approaches have been proposed. For example, Kitaev proposed a method to calculate the Chern number from the reals space in Ref. \cite{KITAEV20062}, which has been applied to quasi-crystalline Chern insulator \cite{PhysRevB.100.214109}.  Another approach utilizes the so-called Bott index which is a real space index for Chern insulators and is used to characterize QHE in disordered systems \cite{loring2011disordered}. Bott index has been applied to two-dimensional quasiperiodic systems \cite{PhysRevX.6.011016,PhysRevB.103.085307,PhysRevB.101.115413,ghadimi2020topological}.

Topological charge pumping in quasiperiodic structures has been experimentally observed in photonic quasicrystals \cite{PhysRevLett.109.106402} and ultra-cold atoms \cite{nakajima2020disorderinduced}. For theoretical studies, charge pumping in the Fibonacci lattice has been studied, for example, by using an interesting connection between Fibonacci lattice and Harper model \cite{PhysRevLett.109.116404} or by approximating quasiperiodic systems with periodic systems with large period \cite{flicker2015,PhysRevResearch.2.042035}. However, a systematic understanding of topological charge pumping in quasiperiodic systems that is based on a general procedure to compute topological index has been still missing.

In this paper, we generalize the Bott index to characterize the charge pumping system and apply it to the quasiperiodic system. The Bott index allows us to compute topological indices of charge pumping regardless of the detail of models. We apply our method to two toy models that are based on the Fibonacci lattice and the Rice-Mele model \cite{PhysRevLett.49.1455} which is a famous model of charge pumping. We demonstrate the topological pumping in these models and study the details of their pumping behaviors.

The rest of this paper is organized as follows. In Sec.~\ref{sec : FibonacciLattice}, we explain the Fibonacci lattice as an example of the quasiperiodic system. In Sec.~\ref{sec : Models}, we introduce two toy models made of the Fibonacci lattice and the Rice-Mele model. In Sec.~\ref{Index}, we briefly review the Bott index and present our generalization of the Bott index for charge pumping. In Sec.~\ref{Results}, we show the result in the RM model and our two models and discuss their unique pumping behaviors. In Sec.~\ref{discussion}, we present a brief discussion.

\section{Fibonacci lattice}\label{sec : FibonacciLattice}
In this paper, we adopt the Fibonacci lattice as a typical example of quasiperiodic systems.
Fibonacci sequence is the sequence of numbers which follows the recursion equation,
\begin{align}
    F_{n+1} = F_{n-1} + F_n,
\end{align}
and the initial conditions $F_0=1,\ F_1 =1$. 
Fibonacci lattice\cite{PhysRevB.40.7413,PhysRevB.93.205153} is obtained by extending this sequence of numbers to a sequence of characters.
It follows the recursion equation $L_{n+1}=L_{n-1}+L_{n}$ and
the initial conditions $L_{0}=\mathbf{a},\ L_{1}=\mathbf{b}$
,where the addition of characters is introduced as $\mathbf{a+b = ab}$.
For example, the first five generations of the Fibonacci lattice are given as
\begin{align*}
   L_0 &= \mathbf{a}, \\
   L_1 &= \mathbf{b}, \\
   L_2 &= \mathbf{ab}, \\
   L_3 &= \mathbf{bab}, \\
   L_4 &= \mathbf{abbab}, \\
   L_5 &= \mathbf{bababbab}.
\end{align*}

This sequence of the Fibonacci lattice can also be obtained by the inflation rule,
\begin{align}
   \mathbf{a} &\xrightarrow{\mathrm{inflation}} \mathbf{b}, \\
   \mathbf{b} &\xrightarrow{\mathrm{inflation}} \mathbf{ab}.
\end{align}
From this rule, we can notice that $\textbf{aa}$ never appears in the Fibonacci lattice. 
As we see in Sec.~\ref{sec: renormalization}, the inflation rule plays an important role in the renormalization group analysis of the quasiperiodic system.

In the limit of $n\rightarrow \infty$, the ratio of $\mathbf{a}$ and $\mathbf{b}$ is described by the golden ratio,
\begin{align}
   \#L_n\ :\  \#\mathbf{a}\ :\ \#\mathbf{b} =& F_n\ :\  F_{n-2}\ :\ F_{n-1} \\
   =& 1 : \tau^2 : \tau,
\end{align}
where $\tau =(\sqrt{5}-1)/2$ denotes the inverse of the golden ratio.

\section{Models}\label{sec : Models}
In this section, we construct two quasiperiodic models based on the Rice-Mele (RM) model~\cite{PhysRevLett.49.1455} and the Fibonacci lattice.
The RM model is a one-dimensional tight-binding model with a staggered potential $h$ and bond alternation $\delta$.
Namely, the Hamiltonian is given as
\begin{align}\label{eq:RMmodel}
   H\left( t \right) =
   \sum_{i=1}^{\infty}&\Big[
      \left\{ 
         \left( \Delta-\left( -1 \right) ^i \delta\left( t \right) \right) c^\dagger_{i+1} c_i +\mathrm{h.c.} \right\} 
         \nonumber
      \\
      & - \left( -1 \right)^i h\left( t \right)   c^\dagger_{i} c_i  
   \Big], \\
   \delta\left( t \right) =& \delta_0\cos\left( 2\pi\frac{t}{T} \right) \ ,\ h\left( t \right) = h_0\sin\left( 2\pi\frac{t}{T} \right),
\end{align}
where $\Delta$ is a uniform component of the nearest-neighbor hopping.
Here, we consider modulating $h$ and $\delta$ in time with a period $T$ to pump charges.
The model has no inversion symmetry when $\Delta,\ \delta$ and $h$ are nonzero. This model is known to show a quantized charge pump characterized by the Chern number $C=1$ as the parameter set $(\delta,h)$ winds around the origin of the parameter space.
\begin{figure}[tb]
   \centering
   \includegraphics[width = \linewidth]{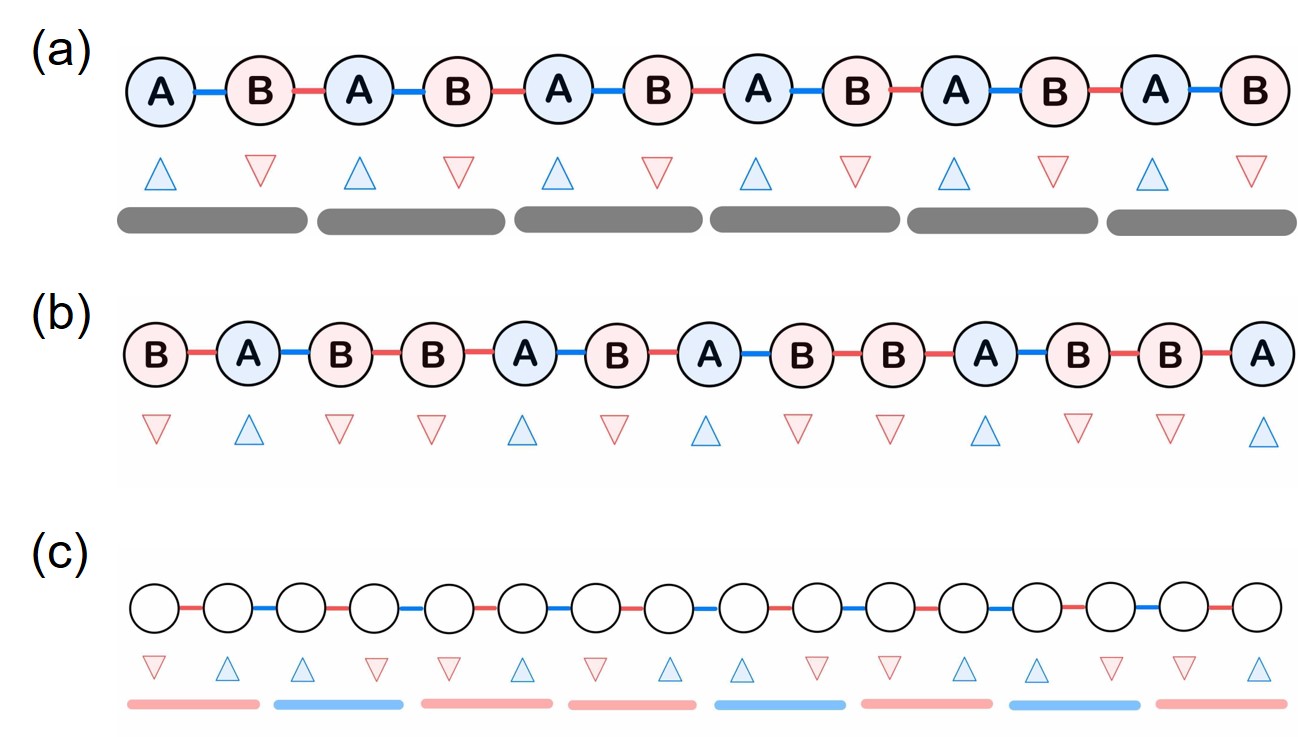}
   \caption{
   Schematic illustration of three models.
   (a) Original Rice-Mele model.
   The blue component is related to $\mathbf{a}$ of the Fibonacci lattice, and red is $\mathbf{b}$.
   Triangles are staggered potentials.
   Black thick lines bars are unit cells.
   (b) Fibonacci-Rice-Mele model at the $6$th generation of the Fibonacci lattice.
   Blue and red lines are following the Fibonacci lattice.
    (c)
    Double-Fibonacci-Rice-Mele model at the $5$th generation of the Fibonacci lattice.
    Red-thick lines and blue-thick lines indicate the blocks which follow the Fibonacci lattice.
    \label{fig : Model}
   } 
\end{figure}

One way to construct a quasiperiodic version of the RM model based on the Fibonacci lattice is to regard the building blocks $\mathbf{a}$ and $\mathbf{b}$ of the Fibonacci sequence ($\mathbf{abbab}\dots$) as the two sublattices (A and B) that constitute the RM model.
While we need to assign $\mathbf{a}$, $\mathbf{b}$ to the bonds as well, here we assign the same character as the site $i$ to 
the bond at the right side of the site $i$, as depicted in Fig.~\ref{fig : Model}(a). Namely, we interpret the factor $-(-1)^i$ in Eq.~(\ref{eq:RMmodel}) as a sign factor depending on the assigned character.
The model corresponding to the character sequence of the Fibonacci lattice, which we call the Fibonacci-Rice-Mele (FRM) model (Fig.~\ref{fig : Model}(b)),
can be obtained by replacing the sign factor as
\begin{align}\label{Ham_FRM}
   H\left( t \right) =\sum_{i=1}^{L} &\left[
   \left\{ 
      \left( \Delta-\left( -1 \right) ^{f_i} \delta\left( t \right) \right) c^\dagger_{i+1} c_i +\mathrm{h.c.} \right\}
   \right.
   \nonumber
   \\
   &\left.- \left( -1 \right)^{f_i} h\left( t \right)   c^\dagger_{i} c_i 
   \right],
\end{align}
where $L$ is the system size and we take lattice constant to be $1$ in this model.
In the case of open boundary conditions, we drop the terms involving $c^\dagger_{L+1} c_L$ and its hermitian conjugate. In the case of periodic boundary conditions, we use the convention $c_{L+1}=c_{1}$.
Here, $f_i$ for the $n$th generation of the Fibonacci lattice is defined as
\begin{align}
   f_i = \begin{cases}
      1  & \mbox{if } i \mbox{th character of } L_n \mbox{ is }\mathbf{a} \\
      0  & \mbox{if } i \mbox{th character of } L_n \mbox{ is }\mathbf{b}
   \end{cases}.
\end{align}

Another quasiperiodic extension of the RM model is based on  translation of the unit cell rather than the sublattice.
The unit cell of the (periodic) RM model can be chosen as either block AB or block BA. By translating the Fibonacci characters $\mathbf{a}$ and $\mathbf{b}$ to these two choices of the unit cell, we can construct another Fibonacci extension of the RM model out of the Fibonacci sequence.  (Namely, the RM model can be represented as $\mathbf{aaa}\dots$ or $\mathbf{bbb}\dots$.)
We call this Fibonacci lattice counterpart of the RM model as
Double-Fibonacci-Rice-Mele (DFRM) model (Fig.~\ref{fig : Model}(c)), whose tight-binding Hamiltonian is given as
\begin{align}\label{Ham_DFRM}
   H\left( t \right) =\sum_{i=1}^{L}&
   \Big[\Big\{ \left( \Delta-(-1)^{f_i}\delta\left( t \right) \right) c^\dagger_{2i} c_{2i-1} \nonumber\\
     +&\left( \Delta+(-1)^{f_i}\delta\left( t \right) \right) c^\dagger_{2i+1} c_{2i}
      +\mathrm{h.c.} \Big\}
   \nonumber \\
   -&(-1)^{f_i}
   \left\{
      h\left( t \right)   c^\dagger_{2i-1} c_{2i-1}
      -h\left( t \right)   c^\dagger_{2i} c_{2i}
   \right\}
   \Big].
\end{align}

\section{Bott index}\label{Index}

In this section, we explain the Bott index, a real space index that characterizes Hall insulators, and generalize the Bott index for application to topological charge pumping in quasiperiodic systems. 

When the momentum $k$ is well defined, 
charge pumping has a conventional characterization by the Chern number that is defined in the momentum space as,
\begin{align}
   C=\frac{1}{2\pi}\sum_{n\in \mathrm{occ.}} \int_{0}^{T} dt \int_\mathrm{BZ} dk \ {F^n_{t,k}},
\end{align}
where, $n$ labels eigenstate below the energy gap, and $F^{n}_{t,k}$ is the Berry curvature.
The subscripts of $t$ and $k$ denote the time and the momentum, respectively.
In contrast, the quasiperiodic systems lacks translational symmetry, where
the momentum becomes ill-defined and
one cannot use Chern numbers to characterize charge pumping.
To avoid this difficulty, here we develop characterization of topological charge pumping based on Bott index that does not rely on the momentum space.

\subsection{A brief overview of Bott index}\label{A brief overview of Bott index}
Bott index is a $K$-theoretic index \cite{loring2011disordered} defined on the real space. 
This index measures the non-commutativity of two operators and can be defined in finite-sized lattice systems.
It is known to be equivalent to the Chern number in the thermodynamic limit (TDL) \cite{loring2011disordered,toniolo2017equivalence}. 
Therefore, in this study, we try to characterize charge pumping in the finite-size system to deduce the behavior in the TDL.

For a rectangular system of $L_x \times L_y $, the Bott index is defined as
\begin{align}
   &I_\mathrm{Bott}=\frac{1}{2\pi}\sum_{i=i} {\mathrm{Im}\log \lambda_i}, \\
   &\left\{
      \lambda_1, \lambda_2, \ldots, \ldots
   \right\}
   =
   \mathrm{Spec}\left\{\hat{V}\hat{U}\hat{V}^\dagger \hat{U}^\dagger\right\},
\end{align}
where $\mathrm{Spec}\{\dots\}$ indicates the set of eigenvalues, and $\hat{U}$ and $\hat{V}$ are the operators defined from Fermi projector $\hat{P}$ and position operators $\hat{x}$ and $\hat{y}$ as follows:
\begin{align}
   & \hat{P} = \sum_{n \in \mathrm{occ.}} \ket{\psi_n}\bra{\psi_n},\\
   &
   \begin{pmatrix}
      \mathbf{0} & \mathbf{0} \\
      \mathbf{0} & \hat{U}
   \end{pmatrix}
   =\hat{P} \exp\left(2\pi i \frac{\hat{x}}{L_x}\right) \hat{P} , \\
   &
   \begin{pmatrix}
      \mathbf{0} & \mathbf{0} \\
      \mathbf{0} & \hat{V}
   \end{pmatrix}
   =\hat{P} \exp\left(2\pi i \frac{\hat{y}}{L_y}\right) \hat{P}.
\end{align}
Here, $n$ is the label of the energy eigenstate $\ket{\psi_n}$, we take the basis of $\hat{P} \exp\left(2\pi i \hat{x}/L_x\right) \hat{P}$ and $\hat{P} \exp\left(2\pi i\hat{y}/L_y\right) \hat{P}$ to be eigenstates $\ket{\psi_n}$.

In the periodic boundary conditions (PBC), the position operators $\hat{x}$ and $\hat{y}$ are ill-defined, because we cannot distinguish $\hat{x}$ and $\hat{x}+L_x$.
In contrast, $\hat{U}$ and $\hat{V}$ can be defined uniquely,
since it is unchanged under the translation by $L_x$ or $L_y$.

\subsection{Physical interpretation of Bott index}
While we explained the mathematical definition of the Bott index above,
the physical interpretation of the index described below is useful for considering the extension to the charge pumping.

Let us assume a one-dimensional periodic lattice with $L$ sites and the lattice constant $1$.
The exponential factor of $\hat{U}^\dagger$ is expressed in the coordinate basis as
\begin{align}\label{eq : Physical interpretation of Bott index original exp}
    \exp\left( -2\pi i \frac{\hat{x}}{L} \right) = 
    \sum_{l=1}^{L}e^{-2\pi i \frac{l}{L}} \ket{l}\bra{l},
\end{align}
where $\ket{l}$ is a state in the site $l$ and the lattice constant is one.
In the periodic system, we can expand the state in the plane wave basis as
\begin{align}\label{eq : Physical interpretation of Bott index plane waves}
    \ket{l} = \frac{1}{\sqrt{L}} \sum_{k_n=2\pi/L}^{2\pi} e^{ik_n l} \ket{k_n},
\end{align}
where $k_n=2\pi n/L$ is the wave number. 
Inserting Eq.(\ref{eq : Physical interpretation of Bott index plane waves}) into Eq.(\ref{eq : Physical interpretation of Bott index original exp}), we obtain the following expression,
\begin{align}
    \exp\left( -2\pi i \frac{\hat{x}}{L} \right) 
    &= \frac{1}{L}\sum_{l}\sum_{k_n,k^\prime_n}
    e^{i(k_n-k^\prime_n-\frac{2\pi}{L})l} \ket{k_n}\bra{k^\prime_n} \\
    &=\sum_{k_n}
    \ket{k_n}\bra{k_n-\frac{2\pi}{L}}.
\end{align}
Therefore, we can understand that $\hat{U}^\dagger$ translates a projected wave function $\hat{P}\ket{\psi}=\ket{\psi_P}$ by $2\pi /L_x$ in the momentum space, and projects it again by $\hat{P}$.

As a consequence, the product of operators $\hat{V}\hat{U}\hat{V}^\dagger \hat{U}^\dagger$ represents a translation of the wave function along the perimeter of the rectangle $[k_x,k_x+2\pi/L_x] \times [k_y,k_y+2\pi/L_y]$ in the $k$-space.
As this path forms a closed loop in the $k$-space, the resulting operator is gauge invariant. Thus the Bott index is given by the sum of the acquired phases over such translation processes.

\subsection{QHE to Charge pumping}
While the Bott index introduced above is defined for two-dimensional systems,
we would like to study the charge pumping in one-dimensional systems in the present study.
To characterize the topological charge pumping, we utilize the interpretation that $\hat{U}$ and $\hat{V}$ are translation operators and convert the Bott index.
As the operator $\hat{V}$ is the translation operator for the $k_y$ direction,
we convert it to the translation operator for the time $t$.
Namely, we replace the translation
\begin{align}
   V^\dagger : k_y \xrightarrow{\mathrm{translation}} k_y +\frac{2\pi}{L_y},
\end{align}
with 
\begin{align}
   V^\dagger : t \xrightarrow{\mathrm{translation}} t+\Delta t,
\end{align}
where $\Delta t$ is a small displacement of $t$.
We should choose $\Delta t$ sufficiently small to reduce the error from the value in the TDL.
This transformation enables us to obtain the Bott index for the charge pumping.

\subsection{Bott index for the charge pumping}\label{sec : Bott index for the charge pumping}
As we have formulated the Bott index for the charge pumping,
let us discuss how the interpretation of the index should be modified.
We show below that the Bott index for the charge pumping can be interpreted as the polarization current.
In the following, we adopt the tight-binding form of the Hamiltonian.

In the two-dimensional system, two directions $x$ and $y$ are coupled together with hopping matrix elements as
\begin{align}
   \hat{H}={\sum_{x^\prime,x}\sum_{y^\prime,y} 
   \left[ 
      H 
   \right] _{x^\prime,y^\prime;x,y} \hat{c}_{x^\prime,y^\prime}^\dagger \hat{c}_{x,y}}.
\end{align}
Hence, it is necessary to diagonalize the entire matrix.
In contrast, in the case of charge pumping in 1D systems, the ``hopping'' matrix elements of different times are zero, and the Hamiltonian is readily in a block diagonal form as
\begin{align}
   \hat{H}
   &=\oplus_t\hat{H}_t
   =\oplus_{t}\sum_{x,x^\prime} \left[ 
      H_t 
   \right] _{x^\prime,x}  \hat{c}_{x^\prime,t}^\dagger \hat{c}_{x,t}.
\end{align}
This means that we just need to diagonalize the block Hamiltonian for each time $t$ to compute the Bott index, which reduces computational cost from $O(L^3T^3)$ to $O(L^3T)$, where $L$ is the lattice size.

Next, we construct the projector $\hat{P}_t$ from the instantaneous Hamiltonian $\hat{H}_t$.
Then, we can rewrite the total projector $\hat{P}$ in the effective two-dimensional system (spanned by $x$ and $t$) into
\begin{align} 
   {\hat{P}=\oplus_{t} \hat{P}_{t}}.
\end{align}
Accordingly, $\hat{U}$ and $\hat{V}$ are also rewritten as
\begin{align}
   \hat{U}&=\oplus_{t} {\hat{P}_{t} \exp\left( 2\pi i \frac{\hat{x}}{L_x} \right) \hat{P}_t}, \\
   \hat{V}&=
   \begin{bmatrix}
      0 & P_{t_1}P_{t_2} &  & \cdots & 0 \\
       & 0 & P_{t_2}P_{t_3} &  &  \\
      \vdots &  &  & \ddots & \vdots \\
       &  &  &  & P_{t_{N-1}}P_{t_{N}} \\
      P_{t_{N}}P_{t_{1}} &  & \cdots &  & 0
   \end{bmatrix}.
\end{align}
Thus the spectrum of $\hat{V}\hat{U}\hat{V}^\dagger \hat{U}^\dagger$ is obtained from 
\begin{align}
   & \hat{V}\hat{U}\hat{V}^\dagger \hat{U}^\dagger \sim \nonumber\\
   & \sum_{t} 
   \hat{P}_{t+\Delta t} \exp\left( 2\pi i \frac{\hat{x}}{L_x} \right)
   \hat{P}_{t+\Delta t} \hat{P}_t
   \exp\left( -2\pi i \frac{\hat{x}}{L_x} \right)\hat{P}_{t},
\end{align}
where $\sim$ means that the spectra are the same at the both sides.
This can be easily seen by writing $P_t$ using the eigenvectors $\left\{ | \psi_i(t) \rangle \right\}$ of $H_t$ as
\begin{align}
   \hat{P}_t =\sum_n {| \psi_n\left( t \right) \rangle \langle \psi_n\left( t \right)  |}.
\end{align}
Using the wave function ${\ket{\psi}}$, we define new matrices $\tilde{U}_{t}$ and $\tilde{V}_{t,t+\Delta t}$ as 
\begin{align}
   \big[ \tilde{U}_t \big]_{n,m}  =&\Big< {\psi_n\left( t \right)  \Big| \exp\left( 2\pi i \frac{\hat{x}}{L_x} \right) \Big| \psi_m \left( t \right)} \Big>, \\
   \big[\tilde{V}_{t ,t+\Delta t}\big]_{n,m}  =&\Big< {\psi_n\left( t\right)  \big| \psi_m \left( t+\Delta t  \right)} \Big>.
\end{align}
Then, we can rewrite $\hat{V}\hat{U}\hat{V}^\dagger \hat{U}^\dagger$ as
\begin{align}
   \hat{V}\hat{U}\hat{V}^\dagger\hat{U}^\dagger
   \sim\sum_{t} 
   \left[ 
      \tilde{V}_{t,t+\Delta t} \tilde{U}_{t+\Delta t} \tilde{V}^\dagger_{t,t+\Delta t} \tilde{U}^\dagger_{t}  
   \right] ,
\end{align}
which leads to the expression of the Bott index $I_\mathrm{Bott}$ for charge pumping as
\begin{align}\label{eq : def_bott_pumping}
   I_\mathrm{Bott} = \sum_{t} 
   \frac{1}{2\pi}\mathrm{Arg}\left[ 
      \tilde{V}_{t,t+\Delta t} \tilde{U}_{t+\Delta t} \tilde{V}^\dagger_{t,t+\Delta t} \tilde{U}^\dagger_{t} 
   \right].
\end{align}
Here, $\mathrm{Arg}[A]$ is a short  hand notation for $\mathrm{Im~Tr~log}[A]$. 

\subsection{Difference from other characterizations}
In this subsection, we compare the present formalism with various approaches to characterize topological pumping adopted in the previous studies \cite{PhysRevB.27.6083,PhysRevA.103.043310,PhysRevB.103.155410,PhysRevResearch.2.042024,PhysRevResearch.2.042035}.



One is the approach based on position expectation value $\langle \hat{x} \rangle$,
\begin{align}
    \big<x\left( t \right)\big> = \bra{\psi(t)} \hat{x}\ket{\psi(t)}.
\end{align}
In the open boundary conditions (OBC), the position operators are well defined,
so we can observe pumping behavior by track the position of the wave function $\ket{\psi(t)}$.

Another is an approach proposed in Ref.\cite{PhysRevLett.82.370}, which can be expressed as
\begin{align}\label{eq : Resta}
   \big<x\left( t \right)\big>= \frac{L_x}{2\pi}\mathrm{Arg}\tilde{U}_{t} .
\end{align}
As we mentioned in Sec.~\ref{A brief overview of Bott index},
the position operators become ill-defined in the periodic system,
but we can define $\tilde{U}_{t}$.
Namely, Eq.(\ref{eq : Resta}) is well defined under the 
periodic boundary condition $\psi(x+L)=\psi(x)$. 

On the basis of Eq.(\ref{eq : Resta}), we can also understand $\tilde{V}_{t,t+\Delta t} \tilde{U}_{t+\Delta t} \tilde{V}^\dagger_{t,t+\Delta t} \tilde{U}^\dagger_{t}  $
as the change of the position expectation value $\big<x\left( t+\Delta t \right)-x\left( t \right)\big>$.
Namely, we can interpret this expression in the limit of $\Delta t \rightarrow 0$, as the polarization current $j(t)$ as
\begin{align}
   j(t) \Delta t =
   \frac{1}{2\pi}
   \mathrm{Arg}\left[ 
      \tilde{V}_{t,t+\Delta t} \tilde{U}_{t+\Delta t} \tilde{V}^\dagger_{t,t+\Delta t} \tilde{U}^\dagger_{t}
   \right],
\end{align}
and the Bott index is expressed as the integral of $j(t)$,
\begin{align}\label{Def_polarization_current}
   I_\mathrm{Bott}= &
   \int_0^T dt\  j\left( t \right).
\end{align}
(Hereafter, we set the charge of an electron $-e$ to be 1 for simplicity.) 
In this formula for $I_\mathrm{Bott}$,
we effectively take the difference of the phases of determinants ($U(1)$ parts) of $\tilde{U}_{t+\Delta t}$ and $\tilde{U}_{t}$ before taking the logarithm.
Therefore, this approach has an advantage that it avoids a jump of the position expectation value coming from the branch cut of Arg.
In addition, as the original Bott index is known to be equivalent to the Chern number in the TDL,
the topological nature of charge pumping in quasiperiodic systems is guaranteed in our Bott index approach. 

\section{\label{Results} Application to the Fibonacci models}

In this section, we apply the index defined in Sec.~\ref{sec : Bott index for the charge pumping} to the three models in Sec. \ref{sec : Models}.

\begin{figure*}[tb]
   \centering
   \includegraphics[width=\linewidth]{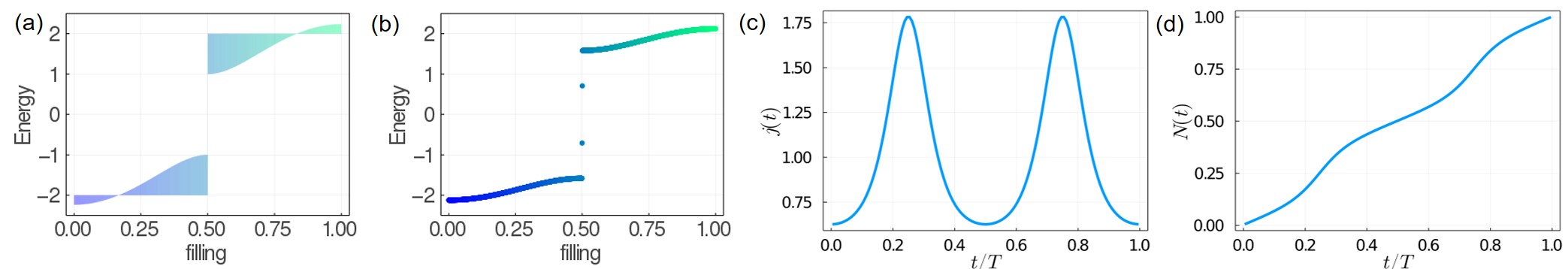}
   \caption{
      Results of numerical calculations in the Rice-Mele model of $1500$ sites.
      (a) Region where energy spectra go through during the pumping under the PBC. The horizontal axis indicates an effective filling factor for the energy level.
      There is no state around $E \in \left[-1 ,1\right]$, during a cycle, which implies that this region
      is a gap in any $t$.
      (b) Energy spectrum at $t = T/4$ under the OBC.
      Two states around the half-filling are the effect of boundary and correspond to edge mode.
      (c,d) Pumping behavior of the half-filled Rice-Mele model of $1500$ sites.
      (c) Polarization current $j\left( t \right) $ as a derivative of the Bott index.
      (d) Pumped charge $N\left( t \right)$ against $t$, which is obtained by accumulating $j(t)$.
      \label{fig : ResultRM}
   }
\end{figure*}

\subsection{Rice-Mele model \label{subsec: RM}}
First, we apply our method to the original RM model.
The energy spectrum of this model is shown in Figs.~\ref{fig : ResultRM}(a) and (b).
In Fig.~\ref{fig : ResultRM}(a), we use the periodic boundary condition (PBC) and the filled region indicates the energy window that are occupied by the $i$th lowest energy level during the cycle. The horizontal axis indicates an effective filling factor $i/N$ with the total number of the states $N$. 
We can observe an energy gap at the half-filling state.
In Fig.~\ref{fig : ResultRM}(b), we plot the instantaneous energy spectrum at $t=T/4$ under the open boundary condition (OBC).
We can find two in-gap levels around the half-filling which are the edge states and are absent in the PBC.

Let us look at the behavior of the Bott index for the half-filling state.
As we have shown in Eq.(\ref{Def_polarization_current}), the Bott index is expressed as the accumulation of the polarization current $j \left(t \right)$,
during one cycle of pumping.
In Fig.~\ref{fig : ResultRM}(c), we show the temporal profile of the polarization current,
which is calculated with $\Delta t = T/128$.
We can observe peaks at $t = T/4 ,3T/4$.
Figure~\ref{fig : ResultRM}(d) shows the pumped charge $N(t)$ defined as
\begin{align}\label{Def : pumped charge}
   N(t) = &
   \int_0^t dt^\prime \  j\left( t^\prime \right),
\end{align}
where we can observe that one particle is pumped during a cycle.
This result agrees with the result from the well-known result for the Chern number $C=1$.

\begin{figure*}[tb]
   \centering
   \includegraphics[width = \linewidth]{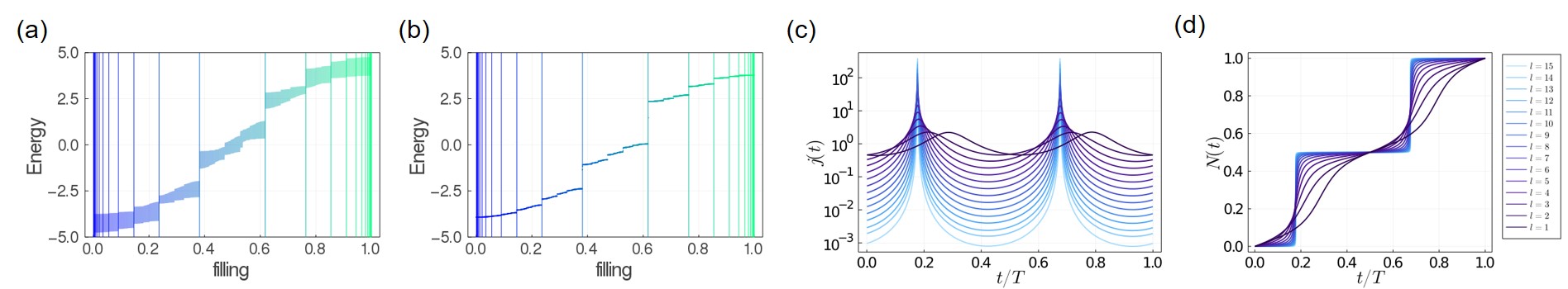}
   \caption{
    Results of numerical calculations in the Fibonacci-Rice-Mele model of the $16$th generation
    (a,b) Energy spectra as a function of the effective filling factor for the energy levels.
    Vertical lines represent fillings related to the power of $\tau$,
    corresponding to the Fibonacci levels.
    (a) Region where energy spectra go through during the pumping under the PBC. The horizontal axis indicates an effective filling factor for the energy level. Gaps at fillings of $\tau$ and $\tau^2$ do not close during a cycle.
    (b) Energy spectrum at $t = T/8$ under the OBC.
    There are also gaps in the Fibonacci levels.
    These gaps are large when they are close to the half-filling state and small when close to filling $0$ and $1$. 
    (c,d) Pumping behavior of the Fibonacci-Rice-Mele model at the fillings of $\tau^l$.
    Data for larger $l$ are represented with lighter colors.
    (c) Polarization current $j\left( t \right) $.
    (d) Pumped charge $N\left( t \right)$ against the time $t$.
    \label{fig : ResultFRM}
   }
\end{figure*}

\subsection{Fibonacci Rice Mele model}
Next, we conduct a similar analysis in the FRM model.
Here we adopt the $16$th generation of the Fibonacci lattice.
As in the original RM model, the topological charge pumping appears when $\Delta > {\left| \delta_0 \right|}, {\left| h_0 \right|}$.
The energy spectrum of this model is shown in Figs.~\ref{fig : ResultFRM}(a) and (b) in a similar manner as in Sec. \ref{subsec: RM}.

As we explained in Sec.~\ref{sec : FibonacciLattice}, for sufficiently large $n$, the relation
\begin{align}
    \frac{F_{n-l}}{F_{n}} \approx \tau^l, \quad (l = 0,1,2,\ldots)
\end{align}
holds.
With this relation, we can specify the ${F_{n-l}}$th lowest eigenstate by the filling factor $\tau^l$, which is independent of the generation $n$.
In the following, we refer to the filling $\tau^l$ as ``Fibonacci levels''.
In Fig.~\ref{fig : ResultFRM}(a), we consider the PBC and filled the region where the energy levels go through during a cycle.
It clearly shows the existence of gaps at the filling factor of $\tau$ and $\tau^2= 1-\tau$ throughout the pumping.
In Fig.~\ref{fig : ResultFRM}(b), we plotted an instantaneous energy spectrum at $t = T/8$ under the OBC. 
This plot suggests the existence of other gaps besides $\tau,\tau^2$.
These gaps are located at fillings of $\tau^3,\tau^4,...$ and $1-\tau^3,1-\tau^4,...$.
This comes from the fractal nature of the Fibonacci lattice.
As we show in appendix \ref{sec: symmetry}, the states at fillings of $\tau^i$ and $1-\tau^i$ are related to each other under the time-reversal and particle-hole symmetry.
Thus we concentrate on the filling $\tau^i$ below.

In Figs.~\ref{fig : ResultFRM}(c) and (d), we show polarization currents and pumped charges at the fillings of $\tau, \tau^2,\ldots, \tau^{15}$.
The deep-blue lines represent the pumping behaviors in the levels with small $l$ 
(i.e. the $F_{n-l}$th energy level with the large Fibonacci number $F_{n-l}$).
We can see that the charge is gradually pumped in this regime.
On the other hand, for the states with larger $l$ (i.e. the $F_{n-l}$th lowest energy state with the smaller Fibonacci number $F_{n-l}$) shown by light blue lines,
the polarization current becomes impulsive and the charge pumping occurs more instantaneously.
In addition, as the power of $\tau$ increases, the time $t$ at which polarization currents become maximum converges to specific values.
We call such behavior of the topological pumping charge pumping that depends on $l$ as ``multi-level topological pumping".

Fibonacci levels are related to each other through inflation and deflation.
As we discuss in Sec.~\ref{sec: renormalization}, by performing real space renormalization group analysis, we can map a state at the filling of $\tau^l$ to another state at the filling of $\tau^{l'}$ ($l'<l$) with modified model parameters.
This implies that the topological charge pumping at the filling of $\tau^l$ is also related to the charge pumping at $\tau^{l'}$ in another model.
Therefore, ``multi-level topological pumping" above can be regarded as a consequence of the fractality of the Fibonacci lattice.

\begin{figure*}[tb]
   \centering
   \includegraphics[width=\linewidth]{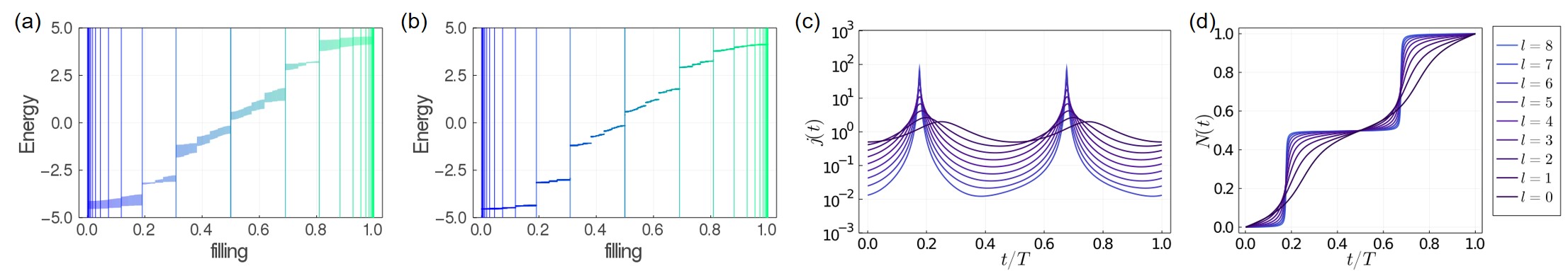}
   \caption{
   Results of numerical calculations in the Double-Fibonacci-Rice-Mele model of the $16$th generation of the Fibonacci lattice.
   (a,b) Energy spectra as a function of the effective filling factor for the energy levels. Vertical lines represents fillings related to the powers of $\tau$.
   (a) Region where energy levels go through during a cycle under the PBC.
   Gaps at the fillings of $\tau/2, \tau^2/2, 1-\tau/2$ and $1-\tau^2/2$  do not close.
   (b) Energy spectrum at $t = T/8$ under the OBC
   (c,d) Pumping behaviors of the Double-Fibonacci-Rice-Mele model at the fillings of $\tau^l/2$.
   (c) Polarization current $j\left( t \right) $. 
   (d) Pumped charge $N\left( t \right)$.
    \label{fig : ResultDFRM}
   }
\end{figure*}

\subsection{Double Fibonacci Rice Mele model}
Finally, we analyze the DFRM model in this subsection.
As in the previous subsection, we adopted the $16$th generation of the Fibonacci lattice.
In Fig.~\ref{fig : ResultDFRM}(a), we filled the region where energy levels go through during the cycle under the PBC.
The instantaneous energy spectrum of this model under the OBC is shown in Fig.~\ref{fig : ResultDFRM}(b), where we can observe many gaps.

As in the FRM model, these gaps are also related to the golden ratio.
These gaps appear at the filling of $\tau/2, \tau^2/2, \ldots$. 
The denominator $2$ comes from the fact that the constituent elements of this model are the blocks consisting of two sublattices (AB or BA) [See Fig.~\ref{fig :  Model}(c)].
In these gaps, we can observe that polarization currents are continuous functions of $t$.

In this model, the polarization current and the Bott index behave as shown in Figs.~\ref{fig : ResultDFRM}(c) and (d).
Up to $l=8$, we can see quantization of charge pumping into as $N(T)=1$.
The plot of $j(t)$ also shows that the particle moves sharply as the power of $\tau$ grows higher and higher.

We note that, in this model, it becomes more difficult to observe quantization of $I_\mathrm{Bott}$ for higher powers of $\tau$, mainly because energy gaps become narrower and the precision of the numerical calculation decreases \cite{toniolo2017equivalence,PhysRevB.98.235425}. 
In order to successfully observe in many levels, it is required to tune the parameters $\Delta,\delta_0,h_0$ more carefully than RM and FRM models.
Such conditions could be understood from detailed real space renormalization group analysis that we explain in Sec. \ref{sec: renormalization}.

\begin{figure}[tb]
   \includegraphics[width=\linewidth]{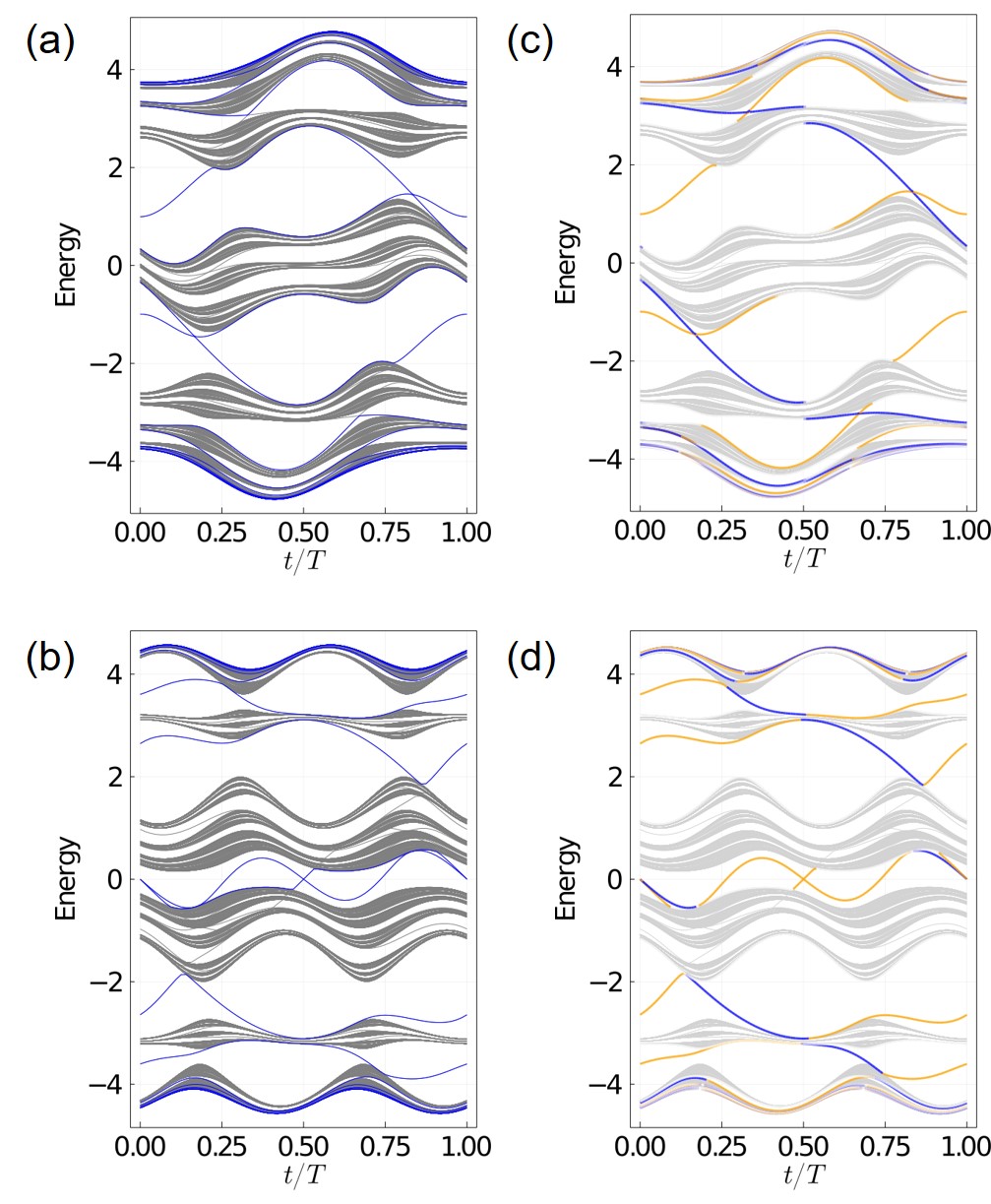}
   \caption{
      Energy spectra against the time $t$ under the OBC.
      (a,c) Energy spectra of the Fibonacci-Rice-Mele model.
      (b,d) Energy spectra of the Double-Fibonacci-Rice-Mele model. 
      (a,b) Fibonacci levels are colored blue and the other levels are colored gray
      (c,d) Position expectation value of the wave functions is shown with color.
      For $f$th level, $0 \leq \langle \hat{x} \rangle / L_x < 1/f$ is blue and $1-1/f < \langle \hat{x} \rangle / L_x \leq 1$ is orange.
      \label{fig:Edge_mode}
   }
\end{figure}

\subsection{Edge mode in the energy spectrum}
Let us discuss the behavior of the edge states, which we showed in Figs.~\ref{fig : ResultFRM}(b) and \ref{fig : ResultDFRM}(b), in more detail.
We plot the energy spectrum in Fig.~\ref{fig:Edge_mode} as a function of pumping parameter $t$.
In Fig.~\ref{fig:Edge_mode}(a) and (b), blue lines for the OBC represent the levels related to the Fibonacci number such as $1$st, $2$nd, $3$rd, $5$th,$\ldots$, $987$th, $1597$th and the levels just above.
The levels related to them under the particle-hole symmetry are also colored blue.
We can clearly see that these states form gapless edge modes in the OBC case.
In addition, there are more edge modes other than Fibonacci levels.
In this paper, we concentrate only on the fillings of $\tau^l$, yet the levels of such as $\tau^l(1-\tau^m)\ (m=1,2,\ldots)$ also related with fractality.
This indicates that there are more topological charge pumpings which come from fractality.

We also calculate the position expectation value $\langle \hat{x} \rangle$ of the lattice,
and color the same energy spectrum following it in Fig.~\ref{fig:Edge_mode}(c) and (d).
Colors for the vertical lines at the Fibonacci levels are determined by following rules.
For Fibonacci numbers $f$ larger than 3, if $\langle \hat{x} \rangle / L_x $ of $f$th level is in $[0,1/f)$, the color is blue, if it is in $(1-1/f,1]$ the color is orange and the other is white.
Accordingly, blue or orange colors of the states traversing the energy gaps indicates  that they are localized to the left or right boundaries, clearly showing that they are edge modes.
This shows that the principle of the bulk-edge correspondence also holds for the topological charge pumping in the quasiperiodic systems.

\begin{figure}[tb]
   \centering
   \includegraphics[width=\linewidth]{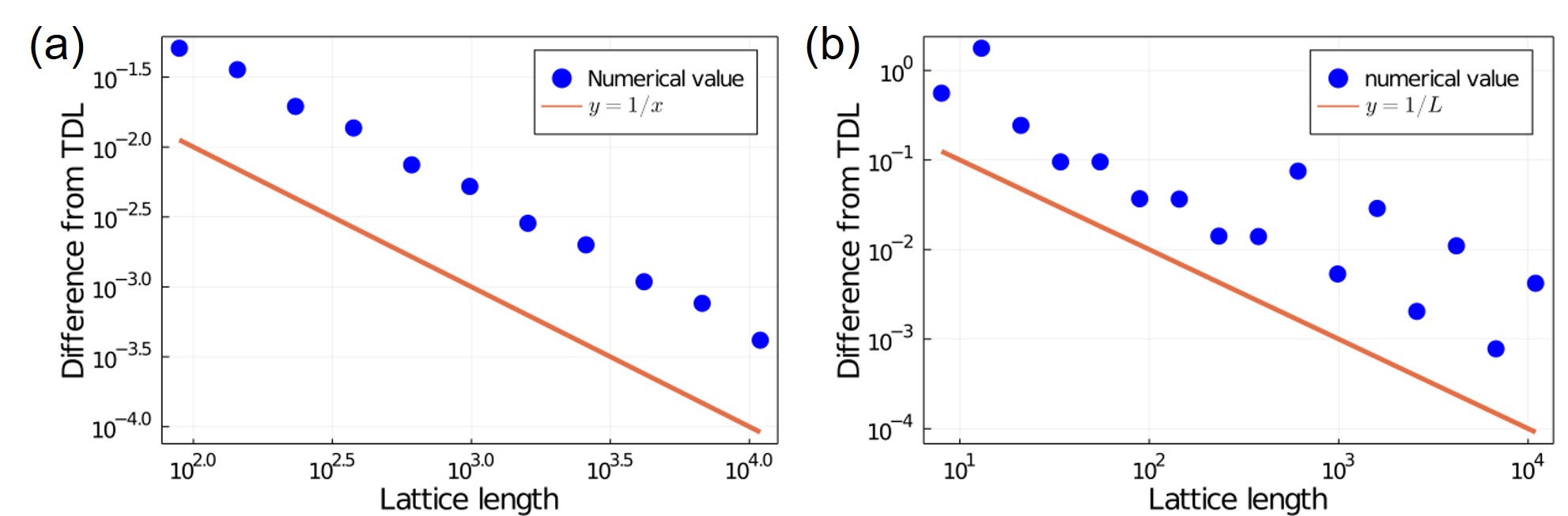}
   \caption{
   The difference of Bott indices from one plotted against the system size for (a) Fibonacci-Rice-Mele model and (b) Double-Fibonacci-Rice-Mele model with the other parameters fixed. 
   The difference is proportional to the inverse of the lattice size. 
   \label{Error_vs_size}
   }
\end{figure}

\subsection{Size dependence of the Bott index}
While the pumping behaviors in the previous subsections are quite close to the ones in the TDL, 
in this subsection, we further check that the values we obtained converge precisely in the TDL.

In Refs.\cite{PhysRevB.98.235425,loring2011disordered}, finite size effect for the case of 2D Chern insulator is already studied by evaluating the difference of the Bott index in the finite-sized system of $L_x\times L_y$ from the quantized value in the TDL,
which shows the deviation scales as  $O((L_xL_y)^{-1})$. 
In the present case, we convert one-direction as the time,
and we translate the wave function by $\Delta t$ using $\tilde{V}$.
Therefore, we can expect the difference from the value in TDL and $\Delta t \rightarrow +0$ to be $O(\Delta t / L)$.

We calculated the Bott indices under the PBC, changing the system size with the other parameters fixed.
The difference of the values from one is plotted in Fig.~\ref{Error_vs_size},
where we can see that the difference is proportional to the inverse of lattice length. This result agrees with the statement of Refs. \cite{PhysRevB.98.235425,loring2011disordered}.
In conclusion, the values we obtained in the finite-sized system converge to one in the TDL.

\subsection{RSRG and ``multi-level-charge pumping''}\label{sec: renormalization}
In this section, we investigate the origin of the ``multi-level topological pumping" and the changes in pumping behaviors.
As explained in Sec.~\ref{sec : FibonacciLattice}, we can change the generation of the Fibonacci lattice through inflation and deflation procedures.
In particular, deflation is an operation to combine a group of sites into a single new site.
It is quite similar to the real space renormalization group (RSRG) studied in Refs.\cite{PhysRevB.42.10329,PhysRevB.85.224205}.
As we explain in Appendix \ref{sec: renormalization_detail},
we can apply the same procedure for the FRM model, at $t = 0$ and $t=T/2$.
At $t=0$, there are three types of renormalization.
One is the renormalization so-called ``atomic renormalization'', which reduces the Fibonacci generation by three.
The others are ``molecular renormalization'' and reduce the generation by two,
where the anti-bonding state and the bonding state appear after renormalization.
As shown in Appendix \ref{sec: renormalization_detail}, at $t=0$, the energy eigenstates in the FRM model of $n$th generation is renormalized into the three groups (bonding, atomic and anti-bonding), where the triplet of parameters, energy shift, $T_w=\Delta-\delta_0,$ and $T_s=\Delta+\delta_0$, is renormalized as
\begin{align}\label{eq: renormalization case A}
   \Big(
         0 ,T_w,  T_s
   \Big) 
   \xrightarrow{\mathrm{deflation}}
   {
   \begin{cases}
      \Big(
         T_s ,
         \frac{\rho}{2}T_w, 
         \frac{\rho}{2}T_s
      \Big) & \mbox{bonding},
      \\
      \Big(
         0 ,
         \rho^2T_w , 
         -\rho^2T_s
      \Big) & \mbox{atomic},
      \\
      \Big(
         -T_s  ,
         \frac{\rho}{2}T_w ,
         -\frac{\rho}{2} T_s 
      \Big) & \mbox{anti-bonding},
   \end{cases}
   }
\end{align}
for each group with $\rho=T_w/T_s$.

We can find three groups in the renormalization process in a fractal structure of the energy spectrum.
Let us consider the renormalization in the FRM model of the $n$th generation.
The group of states that is renormalized into anti-bonding states corresponds to the $F_{n-2}$ lowest energy levels in the energy spectrum due to the negative energy shift.
The group of states that is renormalized into the bonding states corresponds to the $F_{n-2}$ highest energy levels due to the positive energy shift.
Those associated with the atomic states form the energy levels that appear in the middle of the energy spectrum.
As we show in Fig.~\ref{fig:Edge_mode}(b), there exist energy gaps between the states of atomic renormalization and the states of molecular renormalization, and these gaps do not close at any time $t$.
This clearly indicates that the ``multi-level topological pumping'', which take place in these energy gaps, is closely related to the inflation/deflation processes in the Fibonacci lattice.

In summary, the RSRG procedure transforms the original FRM model into other models in lower generations with different parameters, which allows a mapping of the eigenstates to those of the FRM model in a lower generation effectively.
This map also connects the topological charge pumpings at different Fibonacci levels.
This is the origin of the fractal  structure of the energy levels and ``multi-level topological pumping''.

\section{Discussions \label{discussion}}
In this paper, we studied charge pumping in the one-dimensional quasicrystals using the Bott index.
We generalized the Bott index to characterize charge pumping in one-dimensional systems.
In this formulation, the Bott index is directly connected to a time integral of polarization current,
and its computational cost is reduced from $O(L^3T^3)$ to $O(L^3T)$.

By applying our method to the two Fibonacci models, we observed ``multi-level topological pumping".
In both models, there are charge pumpings in the energy levels related to the golden ratio.
This is a result of fractality and bifurcation of the energy spectrum caused by renormalization, inherent to the quasiperiodic crystal structure. This implies that quasicrystals can generally support multilevel topological phenomena with fractality.
Since charge pumping in quasiperiodic structure is already realized in photonic quasicrystals \cite{PhysRevLett.109.106402},
photonic crystals would provide a platform for observing fractality and multi-level topological pumping once Fibonacci sequence is implemented.


Our method for topological pumping based on the Bott index is applicable to other pumping phenomena in quasiperiodic systems. For example, it can be used to study topological spin pumping in spinful electron systems or in topological magnets.  
It can be also applicable to higher dimensional systems where uni-directional topological charge pumping takes place, e.g., realizable in polar heterostructure of two dimensional thin films.

\begin{acknowledgements}
We thank Pasquale Marra and Mikio Furuta
for fruitful discussions.
MY is supported by Forefront physics and mathematics program to drive transformation (FoPM).
This work was partly supported by
JST CREST (JPMJCR19T3) (SK, TM), and JST PRESTO (JPMJPR19L9) (TM). 
\end{acknowledgements}

\appendix
\section{Symmetry of pumping behaviours in the Fibonacci-Rice-Mele model}\label{sec: symmetry}
\begin{figure}[tb]
   \centering
   \includegraphics[width=\linewidth]{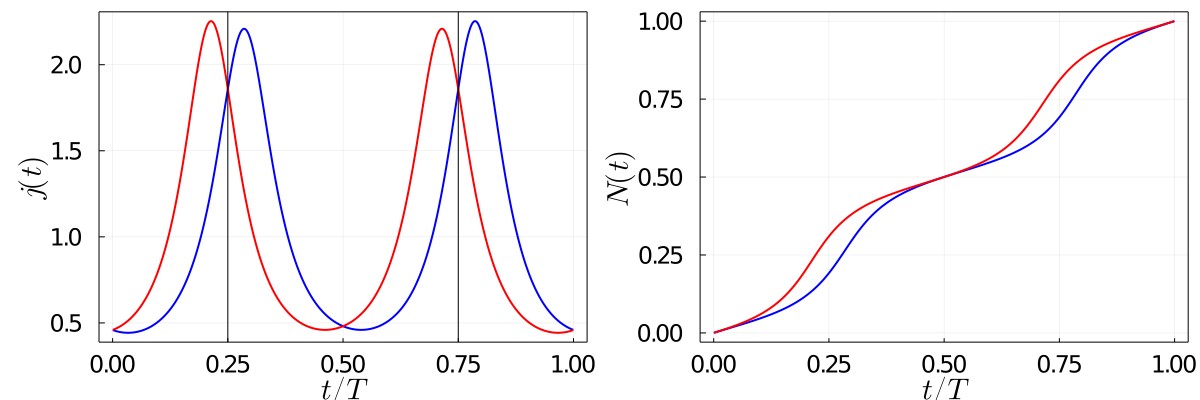}
   \caption{
      Pumping behavior of the Fibonacci-Rice-Mele model at fillings $\tau$ (blue) and $\tau^2$ (red) as a function of $t$.
      \label{fig:FRM_pump_sym}
   }
\end{figure}
As we show in Fig.~\ref{fig:FRM_pump_sym}, the pumping behavior in the filling of $\tau$ and $\tau^2=1-\tau$ are time-reversal (TR) symmetric $(t\rightarrow T-t)$ to each other about $t=T/2$ in the FRM model.
This is the consequence of the TR transformation and particle-hole (PH) transformation as we show below.

First, we demonstrate how the Hamiltonian and the projection operators are transformed.
The FRM model in the present study is written as
\begin{align}\label{eq: sym Ham original}
   H\left( t \right) =
   \sum_{i=1}^{F_n}\Bigg[
   &\Bigg\{ \Big( \Delta-\Big( -1 \Big) ^{f_i}\delta_0\cos\Big( \frac{2\pi}{T}t \Big)\Big)  c^\dagger_{i+1} c_i +\mathrm{h.c.} 
   \Bigg\}
   \nonumber\\
   &- \Big( -1 \Big)^{f_i} h_0\sin\Big( \frac{2\pi}{T}t \Big)   c^\dagger_{i} c_i 
   \Bigg].
\end{align}
For the later convenience, we approximated infinite-sized lattice system by a finite sized lattice system of lattice length is $F_n$. Its time-reversal counterpart ($t\to T-t$) reads
\begin{align}\label{eq: sym Ham after TR}
   H\left( T-t \right) =&
   \sum_{i=1}^{F_n}\Bigg[
   \Big\{ \Big( \Delta-\Big( -1 \Big) ^{f_i}\delta_0\cos\Big( \frac{2\pi}{T}t \Big)\Big)  c^\dagger_{i+1} c_i  \nonumber\\
   &+\mathrm{h.c.} \Big\}
   + \Big( -1 \Big)^{f_i} h_0\sin\Big( \frac{2\pi}{T}t \Big)   c^\dagger_{i} c_i 
   \Bigg].
\end{align}
By further applying PH transformation ($H \to -H$), we obtain
\begin{align}\label{eq: sym Ham after TR}
   -H\left( T-t \right) =&
   \sum_{i=1}^{F_n}\Bigg[
   -\Big\{ \Big( \Delta-\Big( -1 \Big) ^{f_i}\delta_0\cos\Big( \frac{2\pi}{T}t \Big)\Big)  c^\dagger_{i+1} c_i  \nonumber\\
   &+\mathrm{h.c.} \Big\}
   - \Big( -1 \Big)^{f_i} h_0\sin\Big( \frac{2\pi}{T}t \Big)   c^\dagger_{i} c_i 
   \Bigg].
\end{align}
In this Hamiltonian, only the sign of $h$ changes from the original one.
Namely, $H(t)$ and $H$ is related under the unitary transformation
\begin{align}
    H(t) =& - M^\dagger H(T-t) M, \\
    M =& \mathrm{diag}\{1,-1,1,-1,\ldots\}.
\end{align}
Let us calculate the polarization current.
From Eq.(\ref{Def_polarization_current}), polarization current up to $S$th level is defined as follows.
\begin{align}\label{eq : j before TR}
   &j\left( t ; [1,S] \right)  \Delta t=
   \frac{1}{2\pi}\mathrm{Arg}\left[ 
      \tilde{V}_{t,t+\Delta t} \tilde{U}_{t+\Delta t} \tilde{V}^\dagger_{t,t+\Delta t} \tilde{U}^\dagger_{t}  
   \right].
\end{align}
The argument $[a,b]$ means $j$ is calculated using the state from the $a$th lowest eigenstate to the $b$th lowest eigenstate.
Through the TR and PH transformation, the right hand side of Eq.(\ref{eq : j before TR}) is equivalent to
\begin{align}
   \frac{1}{2\pi}\mathrm{Arg}\Big[
      \tilde{V}_{T-t,T-t-\Delta t} \tilde{U}_{T-t-\Delta t}
      \tilde{V}^\dagger_{T-t, T-t-\Delta t } \tilde{U}^\dagger_{T-t}  
   \Big],
\end{align}
once evaluated for PH transformed states.
Specifically, as the $n$th lowest eigenstate of the original Hamiltonian $H(t)$ is transformed into the $n$th highest eigenstate of the $H(T-t)$ by $M$, we take the eigenstates from the highest eigenstate to the $S$th highest eigenstates.
By taking $\Delta t \ll 1$, we obtain
\begin{align}
   &-j\left( T-t ; [F_n-S+1,F_n] \right) \Delta t  \nonumber\\
   &= \frac{1}{2\pi}\mathrm{Arg}\Big[
      \tilde{V}_{T-t,T-t-\Delta t} \tilde{U}_{T-t-\Delta t}
      \tilde{V}^\dagger_{T-t, T-t-\Delta t } \tilde{U}^\dagger_{T-t}  
   \Big].
\end{align}
Thus, we can relate the polarization current before and after the transformation as
\begin{align}
   j\left( t ; [1,S] \right)\Delta t
   =-j\left( T-t ; [F_n-S+1,F_n] \right) \Delta t.
\end{align}
Since the polarization current of the full-filled state is $0$,
i.e.,
$j\left( T-t ; [1,S] \right) +j\left( T-t ; [S+1,F_n] \right) =0$, we obtain
\begin{align}
   j\left( T-t ; [1,F_n-S] \right)
   = -j\left( T-t ; [F_n-S+1,F_n]  \right) .
\end{align}
Therefore, we can also relate the polarization current before and after the transformation as
\begin{align}
   j\left( t ; [1,S] \right)
   = j\left( T-t ; [1,F_n-S]  \right),
\end{align}
As a result, the polarization current up to the $S$th level at the time $t$ and the polarization current up to the $F_n-S$th level of the Hamiltonian at the time $T-t$ are equivalents. 

\section{Details of the real space renormalization group analysis}\label{sec: renormalization_detail}
Here, we explain the renormalization procedure studied in Refs.\cite{PhysRevB.42.10329,PhysRevB.85.224205}.
Before going into the detail of the RSRG, we explain the Brillouin-Wigner perturbation theory.

\subsection{Brillouin-Wigner perturbation theory}
In the following renormalization, we use Brillouin-Wigner perturbation theory (BWPT).
In this appendix, we briefly explain it.
First, we split the original Hamiltonian $H$ into unperturbed term $\hat{H}_0$ and perturbation $\hat{H}_1$.
The eigenequation $E| \psi\rangle = \hat{H}| \psi\rangle$ is rewritten as
\begin{align}
   \left( E-\hat{H}_0 \right)  | \psi\rangle = \hat{H}_1| \psi\rangle.
\end{align}
Using eigenstates of $\hat{H}_0$, we can also construct projector $\hat{P}$,
and split the identity $\mathbf{1}$ into $\hat{P}$ and $\hat{Q}=\mathbf{1}-\hat{P}$.
Rewriting the eigenequation by the projectors, we obtain
\begin{align}
   \left( E-\hat{H}_0 \right)  \hat{P}| \psi\rangle &= \hat{P} \hat{H}_1 \hat{P}| \psi\rangle +\hat{P} \hat{H}_1 \hat{Q}| \psi\rangle, \label{eq: BW_projected}\\
   \left( E-\hat{H}_0 \right)  \hat{Q}| \psi\rangle &= \hat{Q} \hat{H}_1 \hat{P}| \psi\rangle +\hat{Q} \hat{H}_1 \hat{Q}| \psi\rangle.
   \label{eq: BW_unprojected}
\end{align}
To obtain effective Hamiltonian $H_\text{eff}$ for the projected state $|\psi_P\rangle=\hat{P}| \psi\rangle$, 
we express $\hat{Q}| \psi\rangle$ from Eq.(\ref{eq: BW_unprojected}) as
\begin{align}
   \hat{Q}| \psi\rangle &= \frac{\mathbf{1}}{E-\hat{H}_0-\hat{Q}\hat{H}_1\hat{Q}} \hat{Q} \hat{H}_1 \hat{P}| \psi\rangle.
\end{align}
Substituting this into Eq.(\ref{eq: BW_projected}) and multiplying $\hat{P}$ from left, we obtain
\begin{align}\label{eq: BW_eff_ham_exact}
   E|\psi_P\rangle= \left( \hat{P}\hat{H}\hat{P}+\hat{P}\hat{H}_1\hat{Q} \frac{\mathbf{1}}{E-\hat{H}_0-\hat{Q}\hat{H}_1\hat{Q}} \hat{Q} \hat{H}_1 \hat{P} \right) |\psi_P\rangle.
\end{align}
Up to here, there is no approximation.

Expanding $(E-\hat{H}_0-\hat{Q}\hat{H}_1\hat{Q})^{-1}$ using $\left( 1-x \right)^{-1}=1+x+x^2+\cdots $,
Eq.~(\ref{eq: BW_eff_ham_exact}) is deformed into
\begin{align}\label{eq: BW_eff_ham_1storder}
   & E|\psi_P\rangle =\Big( \hat{P}\hat{H}\hat{P}+\hat{P}\hat{H}_1\hat{Q} \frac{\mathbf{1}}{E-\hat{H}_0} \hat{Q} \hat{H}_1 \hat{P}+ \nonumber\\
   &\hat{P}\hat{H}_1\hat{Q} \frac{\mathbf{1}}{E-\hat{H}_0} \hat{Q} \hat{H}_1 \hat{Q} \frac{\mathbf{1}}{E-\hat{H}_0} \hat{Q} \hat{H}_1 \hat{P}+\cdots \Big) |\psi_P\rangle.
\end{align}
It is hard to know $E$ as it is the eigenvalue of the original Hamiltonian $\hat{H}$.
In the following, we approximate it by the eigenvalues of $\hat{P}\hat{H}_0\hat{P}$ and $\left( 1-x \right)^{-1}$ by $1+x+x^2$.
In conclusion, we obtain an approximated effective Hamiltonian
\begin{align}\label{eq: BW_eff_ham_2ndorder}
   &H_\text{eff}=\Big( \hat{P}H\hat{P}+\hat{P}\hat{H}_1\hat{Q} \frac{\mathbf{1}}{E-\hat{H}_0} \hat{Q} \hat{H}_1 \hat{P}+ \nonumber\\
   &\hat{P}\hat{H}_1\hat{Q} \frac{\mathbf{1}}{E-\hat{H}_0} \hat{Q} \hat{H}_1 \hat{Q} \frac{\mathbf{1}}{E-\hat{H}_0} Q \hat{H}_1 \hat{P} \Big).
\end{align}

\subsection{Detail of the renormalization procedure}\label{appendix: Renormalization}
\begin{figure}[tb]
   \centering
   \includegraphics[width=0.9\linewidth]{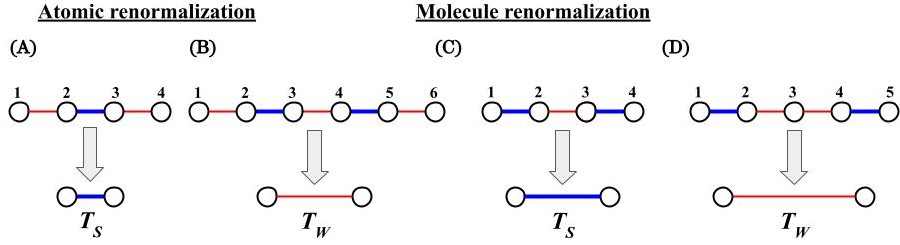}
   \caption{RG transformation in the ``Case A'' of \cite{PhysRevB.42.10329}.
   The blue thick lines represent bonds related to $\mathbf{a}$ of the Fibonacci lattice and the red thin lines represent bonds of $\mathbf{b}$.
   \label{fig : CaseA}
   }
\end{figure}
Here, we derive Eq.~(\ref{eq: renormalization case A})
and explain the renormalization procedure in detail.
The outline of the calculation is as follows.
First, we split the Hamiltonian into the unperturbed term $\hat{H}_0$ and the perturbation $\hat{H}_1$.
From $\hat{H}_0$, we calculate eigenvalues and corresponding eigenstates. 
After that, we construct the effective Hamiltonian $H_\text{eff}$ using BWPT.
By sandwiching $H_\text{eff}$ by the eigenstates of $\hat{H}_0$, we obtain renormalized values of couplings and staggered potentials.

In Ref.\cite{PhysRevB.42.10329}, two models called the diagonal model and the off-diagonal model are studied.
The diagonal model corresponds to the FRM model at $t=0$ [$\delta(0)=\delta_0,\ h(0)=0$] and $t=T/2$ [$\delta(0)=-\delta_0,\
h(0)=0$], while the off-diagonal model corresponds to the FRM model at $t= T/4$ [$\delta(0)=0,\ h(0)=h_0$] and $t=3T/4$ [$\delta(0)=0, h(0)=-h_0$].

In this appendix, we focus only on the ``Case A'' of Ref. \cite{PhysRevB.42.10329}, which corresponds to the case of $t=0$ in the FRM model.
For convenience, we label the sites $| 1 \rangle, | 2 \rangle \ldots $ from left as shown in Fig.~\ref{fig : CaseA}.
At $t=0$, $\delta(0) = \delta_0$ and $h(0)=0$.
Namely, there are no on-site potentials, and only bond modulation $\delta$ changes according to the Fibonacci sequence.
We express bonds for $\mathbf{a}$ as $T_s = \Delta+\delta_0$ and
bonds for $\mathbf{b}$ as $T_w = \Delta-\delta_0$.
In the following, we take $\Delta,\,\delta_0 > 0$ and treat $T_w$ as a perturbation.
Then, the unperturbed Hamiltonian $H_0$ becomes block-diagonal, as only the bonds with $\mathbf{a}$ have nonzero matrix elements.
Since $\mathbf{aa}$ does not appear in the Fibonacci sequence [See Sec.~\ref{sec : FibonacciLattice}], the unperturbed Hamiltonian is composed of $1\times 1$ block matrices $0$ and $2\times 2$ block matrices,
\begin{align}
   \begin{pmatrix}
      0 & T_s \\
      T_s & 0 
   \end{pmatrix}.
\end{align}
The eigenvalues and eigenvectors of this matrix are $\pm T_s$ and $\ket{\psi_{\pm}} = \frac{1}{\sqrt{2}}\left( | 1\rangle \pm | 2\rangle \right) $.
Here, we call $\ket{\psi_+}$ the bonding state and $\ket{\psi_-}$ the anti-bonding state.
We refer to these states lying on two sites connected to each other with $T_s$ as the molecular states. 
We call the eigenstates of $0$ block matrices, which is localized to a site not connected to other sites, the atomic states.

As we see in the following, we can obtain two different types of renomalizations, the atomic renormalization and the molecular renormalization, according to the choice of the projector $\hat{P}$.

\subsubsection{Atomic renormalization\label{sec : case A atomic}}
In the atomic renormalization, we renormalize bonds between atomic sites into new bonds.
The projector $P$ for this renormalization is 
\begin{align}
    P = \sum_{i\in \text{atomic}} \ket{i}\bra{i},
\end{align}
where the summation is taken over all atomic sites of $H_0$, and $\ket{i}$ is an atomic state of $H_0$. 
This renormalization is illustrated in Figs.~\ref{fig : CaseA}(A) and (B).

In the atomic renormalization, we have two types of bond configurations, and the other configurations are prohibited by the inflation rule. 
Specifically, the configuration which has three molecule states between two atomic sites($\mathbf{bababab}$) is prohibited, as it does not follow the inflation rule, if we deflate it twice.
The first is $\mathbf{bab}$ shown in Fig.~\ref{fig : CaseA}(A). 
In this case, the unperturbed Hamiltonian is 
\begin{align}
   H_0
   =\begin{pmatrix}
      0 & 0 & 0 & 0 \\
      0 & 0 & T_s & 0 \\
      0 & T_s & 0 & 0 \\
      0 & 0 & 0 & 0
   \end{pmatrix}.
\end{align} 
Eigenvectors of $H_0$ are $| 1 \rangle ,\frac{1}{\sqrt{2}}\left( | 2 \rangle \pm | 3 \rangle \right), | 4 \rangle$.
$H_1$ is the remaining components $T_w$,
\begin{align}
   H_1
   =\begin{pmatrix}
      0 & T_w & 0 & 0 \\
      T_w & 0 & 0 & 0 \\
      0 & 0 & 0 & T_w \\
      0 & 0 & T_w & 0
   \end{pmatrix}.
\end{align} 
By constructing an approximated effective Hamiltonian $H_\text{eff}$ using Eq.(\ref{eq: BW_eff_ham_1storder}), we can calculate effective coupling as
\begin{align}
   \langle 1 | H_\text{eff} | 4 \rangle
   = -\frac{T_w^2}{T_s} = -\rho^2T_s.
\end{align}
Here, $\rho=T_w/T_s$.

The next is $\mathbf{babab}$ shown in Fig.~\ref{fig : CaseA}(B).
The unperturbed Hamiltonian is expressed as
\begin{align}
   H_0
   =\begin{pmatrix}
      0 & 0 & 0 & 0 & 0 & 0 \\
      0 & 0 & T_s & 0 & 0 & 0 \\
      0 & T_s & 0 & 0 & 0 & 0 \\
      0 & 0 & 0 & 0 & T_s & 0 \\
      0 & 0 & 0 & T_s & 0 & 0 \\
      0 & 0 & 0 & 0 & 0 & 0 
   \end{pmatrix},
\end{align} 
while the perturbation is given as
\begin{align}
   H_1
   =\begin{pmatrix}
      0 & T_w & 0 & 0 & 0 & 0 \\
      T_w & 0 & 0 & 0 & 0 & 0 \\
      0 & 0 & 0 & T_w & 0 & 0 \\
      0 & 0 & T_w & 0 & 0 & 0 \\
      0 & 0 & 0 & 0 & 0 & T_w \\
      0 & 0 & 0 & 0 & T_w & 0 
   \end{pmatrix}.
\end{align} 
By constructing an approximated effective Hamiltonian $H_\text{eff}$ using Eq.(\ref{eq: BW_eff_ham_1storder}), we can calculate effective coupling as
\begin{align}
   \langle 1 | H_\text{eff} | 6 \rangle
   =\frac{T_w^3}{T_s^2}= \rho^2 T_w.
\end{align}

Here, let us consider how the generation of the Fibonacci lattice changes.
From the two calculations above, groups of $\mathbf{bab}$ are renormalized to stronger couplings $-\rho^2 T_s$, and groups of $\mathbf{babab}$ are renormalized to weaker couplings $\rho^2 T_w$.
Groups of three bonds are renormalized to bonds $\mathbf{a}$ and groups of five bonds are renormalized to bonds $\mathbf{b}$.
These correspond to $F_3= 3 \rightarrow F_0=1$ and $F_4= 5 \rightarrow F_1=1$,
so the generation is reduced by three.

\subsubsection{Molecular renormalization\label{sec : case A molecule}}
In the molecular renormalization, we renormalize the hopping from one molecular state to another molecular state into a new hopping.
Since we have bonding state $| \psi_{+} \rangle$ and anti-bonding state $| \psi_{-} \rangle$, we can choose whether we renormalize to the bonding state or anti-bonding state. 
Namely, we have two choices for the projector,
\begin{align}
    P_{\pm} = \sum_{i\in\text{molecular}} \ket{\psi_{\pm},i}\bra{\psi_{\pm},i},
\end{align}
where $\ket{\psi_\pm,i}=\frac{1}{\sqrt{2}}(\ket{i}\pm\ket{i+1})$ denotes the bonding $(+)$ and anti-bonding $(-)$ state, and the summation is taken over all (anti-)bonding sites of $H_0$.

The first case is $\mathbf{aba}$ shown in Fig.~\ref{fig : CaseA}(C).
The unperturbed Hamiltonian is
\begin{align}
   H_0=
   \begin{pmatrix}
      0 & T_s & 0 & 0 \\
      T_s & 0 & 0 & 0\\
      0 & 0 & 0 & T_s \\
      0 & 0 & T_s & 0
   \end{pmatrix},
\end{align}
whose eigenvectors are 
$| \psi_{\pm L}\rangle = \frac{1}{\sqrt{2}}\left( | 1 \rangle \pm | 2 \rangle  \right) $ and
$| \psi_{\pm R}\rangle = \frac{1}{\sqrt{2}}\left( | 3 \rangle \pm | 4 \rangle  \right) $.
The remaining component of the original Hamiltonian is the perturbation
\begin{align}
   H_1=
   \begin{pmatrix}
      0 & 0 & 0 & 0 \\
      0 & 0 & T_w & 0\\
      0 & T_w & 0 & 0 \\
      0 & 0 & 0 & 0
   \end{pmatrix}.
\end{align}
By constructing an approximated effective Hamiltonian $H_\text{eff}$ using Eq.~(\ref{eq: BW_eff_ham_1storder}), we can calculate effective coupling as
\begin{align}
   \langle \psi_{\pm L} | H_\text{eff} | \psi_{\pm R} \rangle
   =\pm \frac{\rho}{2} T_s.
\end{align}

The next is $\mathbf{abba}$ shown in Fig.~\ref{fig : CaseA}(D).
In this case, the unperturbed Hamiltonian $H_0$ is
\begin{align}
   H_0=
   \begin{pmatrix}
      0 & T_s & 0 & 0 & 0 \\
      T_s & 0 & 0 & 0 & 0 \\
      0 & 0 & 0 & 0 & 0 \\
      0 & 0 & 0 & 0 & T_s \\
      0 & 0 & 0 & T_s & 0 
   \end{pmatrix},
\end{align}
and eigenvectors are 
$| \psi_{\pm L}\rangle = \frac{1}{\sqrt{2}}\left( | 1 \rangle \pm | 2 \rangle  \right) $ , 
$| \psi_0\rangle = | 3\rangle$and 
$| \psi_{\pm R}\rangle = \frac{1}{\sqrt{2}}\left( | 4 \rangle \pm | 5 \rangle  \right) $.
The perturbation $H_1$ is
\begin{align}
   H_1=
   \begin{pmatrix}
      0 & 0 & 0 & 0 & 0 \\
      0 & 0 & T_w & 0 & 0 \\
      0 & T_w & 0 & T_w & 0 \\
      0 & 0 & T_w & 0 & 0 \\
      0 & 0 & 0 & 0 & 0 
   \end{pmatrix}.
\end{align}
By constructing an approximated effective Hamiltonian $H_\text{eff}$ using Eq.(\ref{eq: BW_eff_ham_1storder}), we can calculate effective coupling as
\begin{align}
   \langle \psi_{\pm L} | H_\text{eff} | \psi_{\pm R} \rangle
   = \frac{\rho}{2}T_w.
\end{align}
In this molecular renormalization, the left-most bond and the right-most bond are shared with the neighboring groups.
Hence the size of groups depicted in Fig.~\ref{fig : CaseA}(C) is $F_2 =2$ and Fig.~\ref{fig : CaseA}(D) is $F_3 = 3$.
Therefore, the generation is reduced by two.

In summary, the triplet of parameters, energy shift, $T_w$ and $T_s$, of the FRM model at $t=0$ is renormalized as
\begin{align}
   \Big(
         0 ,T_w,  T_s
   \Big) 
   \xrightarrow{\mathrm{deflation}}
   {
   \begin{cases}
      \Big(
         T_s ,
         \frac{\rho}{2}T_w, 
         \frac{\rho}{2}T_s
      \Big) & \mbox{bonding},
      \\
      \Big(
         0 ,
         \rho^2T_w , 
         -\rho^2T_s
      \Big) & \mbox{atomic},
      \\
      \Big(
         -T_s  ,
         \frac{\rho}{2}T_w ,
         -\frac{\rho}{2} T_s 
      \Big) & \mbox{anti-bonding}.
   \end{cases}
   }
\end{align}
The $F_n$ eigenstates of the FRM model of $n$th generation are divided into $F_{n-2}$ bonding states, $F_{n-3}$ atomic states and $F_{n-2}$ anti-bonding states, which is consistent with $F_n=2F_{n-2}+F_{n-3}$.

\bibliography{ref}

\end{document}